# Measure dependence of 2D simplicial quantum gravity *

Christian Holm[a] and Wolfhard Janke[ab]

[a]Institut für Theoretische Physik, Freie Universität Berlin, 14195 Berlin, Germany

[b]Institut für Physik, Johannes Gutenberg-Universität Mainz, 55099 Mainz, Germany

We study pure 2D Euclidean quantum gravity with $R^2$ interaction on spherical lattices, employing Regge's formulation. We attempt to measure the string susceptibility exponent $\gamma_{\rm str}$ by using a finite-size scaling Ansatz in the expectation value of $R^2$. To check on effects of the path integral measure we investigate two scale invariant measures, the "computer" measure $dl/l$ and the Misner measure $dl/\sqrt{A}$.

## 1. INTRODUCTION

Along with the theoretical achievements of string theories 2D Euclidean quantum gravity has become a rather well understood subject. In contrast to the classical theory, which is dynamically trivial, the quantum theory possesses a rather rich structure. One of the interesting aspects is the intrinsic fractal structure of space time which shows up in the divergence of the partition function $Z(A)$ with increasing area $A$, which is governed by the string susceptibility exponent $\gamma_{\rm str}$:

$$Z(A) \propto A^{\gamma_{\rm str}-3} e^{-\lambda_R A}, \qquad (1)$$

where $\lambda_R$ denotes the renormalized cosmological constant. The exponent $\gamma_{\rm str}$, which depends on the genus $g$ of the surface, has been calculated by conformal field theory methods [1] to be

$$\gamma_{\rm str} = 2 - \frac{5}{2}(1-g). \qquad (2)$$

Because $\gamma_{\rm str} = 2$ gives the classical result quantum effects can only be seen for $g \neq 1$. For a spherical surface the prediction $\gamma_{\rm str} = -1/2$ was confirmed by matrix model methods, which gave rise to the dynamically triangulated surface approach to quantum gravity [2].

For the alternative approach using Regge calculus [3] the situation is less clear. Using the $dl/l$ measure, Gross and Hamber [4] found weak numerical evidence for $\gamma_{\rm str} = -1/2$, while Bock and Vink [5] clearly did not see the dependence of (2). We therefore reinvestigate this problem using a constrained MC approach close to Ref. [4], but with a spherical lattice topology as used in Ref. [5]. Furthermore we want to investigate the measure dependence of the Regge approach by employing the Misner measure in addition to the commonly used $dl/l$ measure.

## 2. SIMULATION

We simulated the partition function

$$Z = \int \mathcal{D}\mu(l) \exp\,(-\sum_i (\lambda A_i + a\frac{\delta_i^2}{A_i})) \qquad (3)$$

where $\delta_i = 2\pi - \sum_{t\supset i} \theta_i(t)$ is the deficit angle with $\theta_i(t)$ being the dihedral angle at vertex $i$ and $A_i = \sum_{t\supset i} A_t/3$ denotes the baricentric area with $A_t$ being the area of a triangle $t$. The notation is identical to that used in Ref. [6]. Because we want to study a constant area ensemble, we held the area fixed to its initial value $A = \sum_i A_i = N_2/2$ during the update, with $N_2$ denoting the number of triangles, so that the only dynamical term is the $R^2$-interaction. As global lattice topology we used the surface of a three-dimensional cube, see Ref. [5].

We studied two scale invariant lattice measures. One is the most commonly used "computer" measure $\mathcal{D}\mu(l) = \prod_{\langle ij\rangle} dl_{ij}/l_{ij}$, which has no continuum analogue, while the other one is the Misner measure $\mathcal{D}\mu(l) = \prod_{\langle ij\rangle} dl_{ij}/\sqrt{A}_{ij}$, which is the lattice version of the continuum measure

*Work supported in part by the EEC under contract No. ERBCHRX CT93043.

$\mathcal{D}\mu[g] = \prod_x g^{-3/2} \prod_{\mu \geq \nu} dg_{\mu\nu}$ [4]. Here we defined $A_{ij} = \sum_{t \supset \langle ij \rangle} A_t/3$ as the area associated with a link $\langle ij \rangle$.

To update the links we used a standard multi-hit Metropolis update with a hit rate ranging from 1 ... 3. In addition to the usual Metropolis procedure a change in link length is only accepted, if the links of a triangle fulfill the triangle inequalities, and for the Misner measure we also required the fatness [7] condition $f \geq 0.5$. The fatness $f$ is defined as $f = A_t/l_{\max}^2$, with $l_{\max} = \max\{l_i \subset A_t\}$. We ran simulations for both measures at values of $a = 1.25$ and $a = 10.0$, and for the $dl/l$ measure we performed additional runs at $a = 5.0$. The size of the lattices varied from 218 up to 9128 lattice sites, corresponding to 648 up to 27378 link degrees of freedom. For each run we recorded about 50000 measurements of the curvature square $R^2 = (\sum_i \delta_i^2/A_i)/N_0$, where $N_0$ denotes the number of vertices. We used the standard reweighting method to obtain additional data close to the simulated value of $a$. The statistical errors were computed using standard jack-knife errors on the basis of 20 blocks.

## 3. RESULTS

The finite-size scaling (FSS) of $\langle R^2 \rangle$ is expected [5] to be of the form

$$a\langle R^2 \rangle = c_0 + c_1/N_0 + \ldots . \tag{4}$$

The coefficient $c_1$ should depend on the dimensionless parameter $\tilde{a} = a/(A/N_2)$ via $c_1(\tilde{a}) = \gamma_{\rm str}^{\rm eff} - 2$. In the limit that $\tilde{a} \to \infty$ it is expected that $\gamma_{\rm str}^{\rm eff} \to \gamma_{\rm str}$.

### 3.1. $dl/l$ measure

For the computer measure we did not encounter any equilibration problems. The runs behaved well under different starting conditions, and the autocorrelation times of $R^2$ turned out to be about 5 measurements for all lattice sizes. In Fig. 1 we have plotted $a\langle R^2 \rangle$ versus $1/N_0$ for all simulated values of $a$. The scaling Ansatz (4) works well for $a = 1.25$, but shows small deviations for smaller lattices and larger $a$-values. From linear least square fits we obtain $c_1(2.5) = 2.11(5)$, $c_1(10) = 4.0(1)$, and $c_1(20) = 6.2(1)$ respectively.

Figure 1. Scaling of $a\langle R^2 \rangle$ for the $dl/l$ measure.

These values all give clearly a positive $\gamma_{\rm str}^{\rm eff}$. This is in qualitative agreement with the findings of Ref. [5], and in disagreement with (2). To obtain additional data for the extrapolation to large $\tilde{a}$ we tried to reweight our time series in $a$, but the accessible reweighting range turned out to be rather small ($\pm 0.25$). Nevertheless we show in Fig. 2 our simulation points together with 5 additional values obtained by reweighting. On the basis of just three measured points one observes an almost linear dependence. This differs considerably from the behaviour found in Ref. [5]. To some extent this can be traced back to the fact that we used $N_0$ instead of $N_2$ in the scaling Ansatz (4). This result could mean two things, one is that the Ansatz (4) is not valid, and one simply cannot extract in this way a finite $\gamma_{\rm str}$. The other is, that Regge calculus with the $dl/l$ measure cannot reproduce the prediction (2).

Figure 2. $c_1$ as a function of $\tilde{a}$.

Figure 3. $a\langle R^2\rangle$ vs. $1/N_0$ for the Misner measure.

### 3.2. Misner measure

To investigate the measure dependence of the results of the previous section, we performed for $a = 1.25$ and $a = 10$ simulations with the Misner measure. As already in our study of Ising spins coupled to gravity [8], we observed a qualitatively *completely* different behaviour. The autocorrelation times of $R^2$ became extremely long, and we ran into equilibration problems. The outcome of our simulations can be inspected in Fig. 3. We cannot observe a clear scaling behaviour. For the good-willing observer it looks as if the larger lattices show a crossover to negative values of $\gamma_{\rm str}^{\rm eff}$.

## 4. CONCLUSIONS

We tried to measure the string susceptibility $\gamma_{\rm str}$ using a FSS Ansatz in $R^2$. The results we obtained for the $dl/l$ measure make it appear likely that the Ansatz (4) does not work, because $\gamma_{\rm str}^{\rm eff}$ seems to scale linearly with $\tilde{a}$. If there is an upper bound at all, then it would be clearly positive, excluding the prediction of (2). For a different approach to measure $\gamma_{\rm str}$ see Ref. [9], where it was claimed on the basis of a numerical study of the universal loop length distribution, that the $dl/l$ measure seems to give the correct distribution, but only on very large lattices.

For the Misner measure we can only state, that the qualitative behaviour of the model is completely different, which is also true, if one includes matter fields [8]. The numerical difficulties one encounters make a quantitative statement at the present status impossible. It looks as if the larger lattices are better behaved due to some sort of self-averaging, and that there is a possible crossover to negative $\gamma_{\rm str}^{\rm eff}$. Perhaps the Regge approach with its fixed incidence matrix can simulate a fractal structure only on large lattices by distorting the local link lengths rather vividly, as seems to happen with the Misner measure. The DTS approach can mimic this behaviour due to its fluctuating local topology much easier.

## ACKNOWLEDGEMENTS

We thank W. Bock for useful discussions. W.J. gratefully acknowledges a Heisenberg fellowship by the DFG. The numerical simulations were performed on the North German Vector Cluster (NVV) under grant bvpf01.

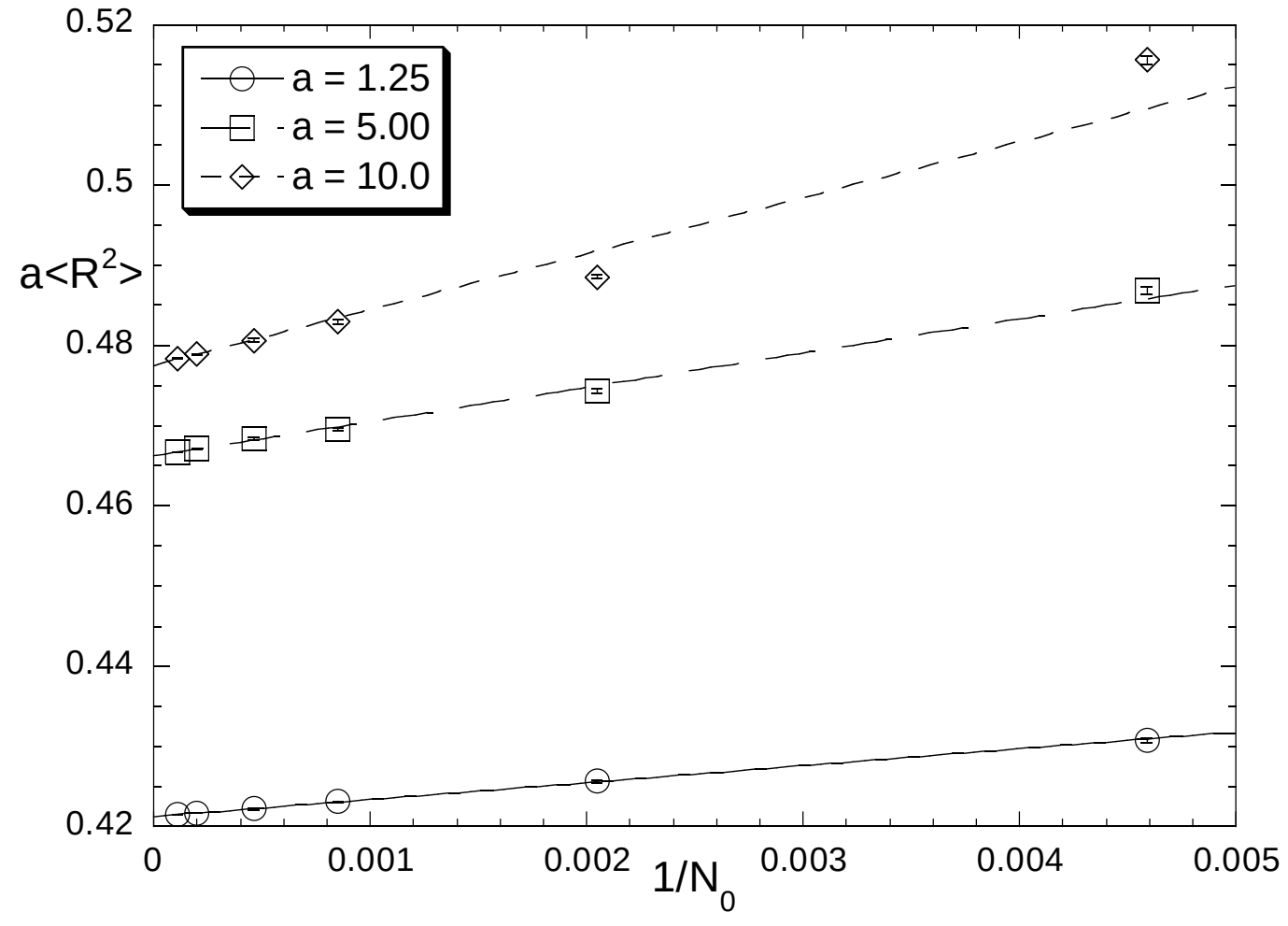

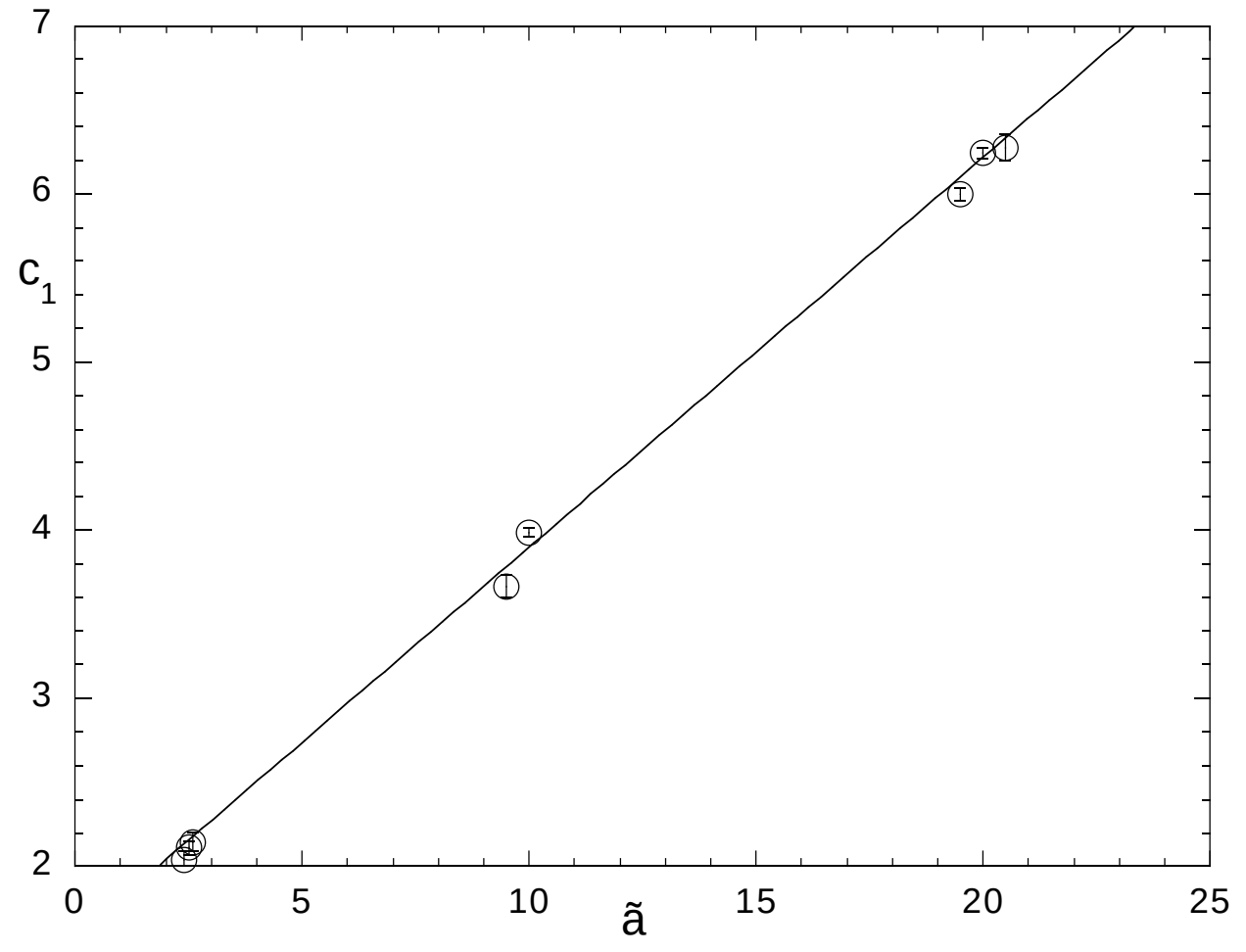

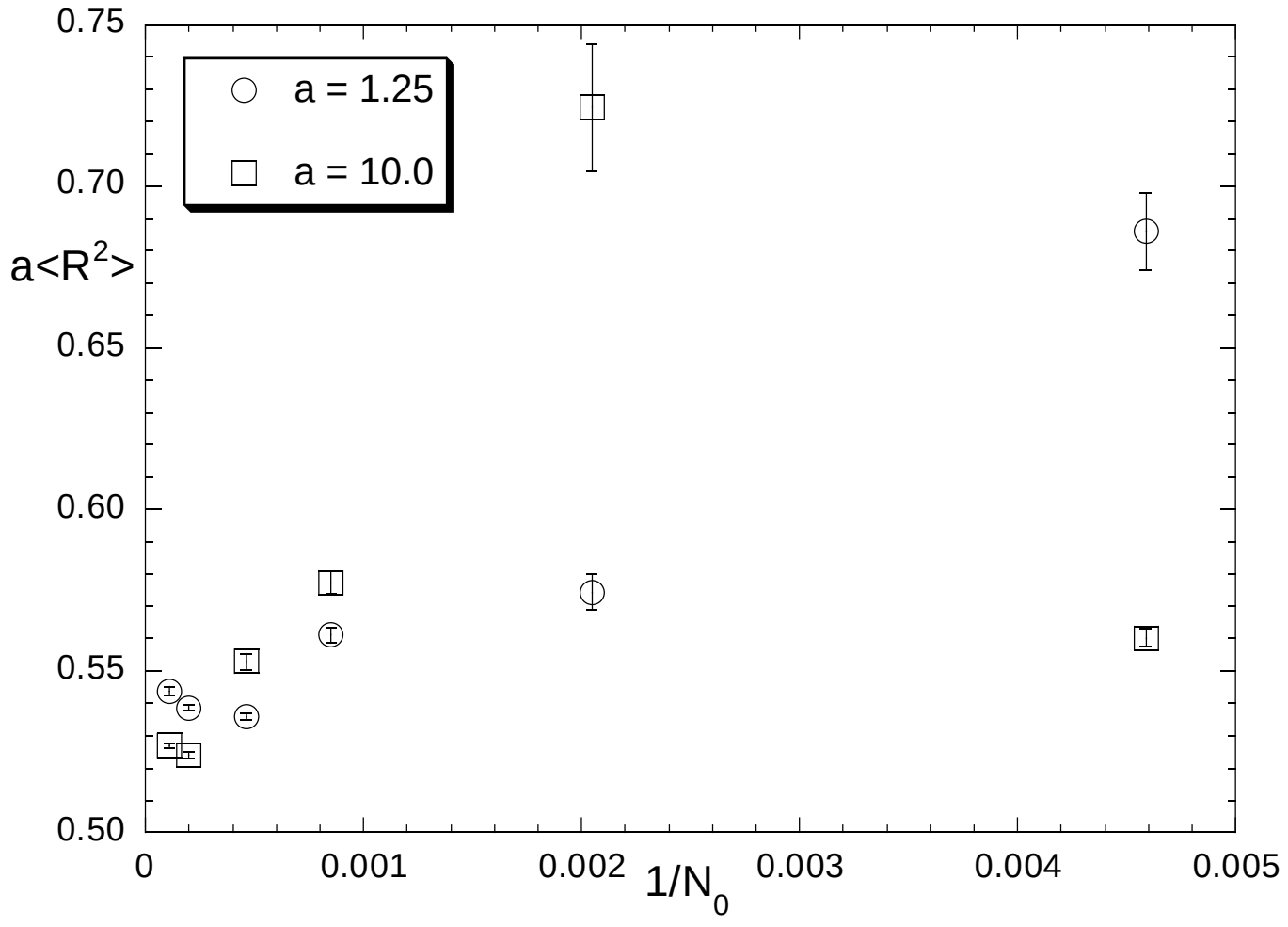